Reduction of temperature drift in refractive-index-sensing optical frequency comb by active-dummy compensation of dual-comb configuration

Shogo Miyamura[1,*], Masayuki Higaki[2,*], Shuji Taue[3], Yoshiaki Nakajima[4], Yu Tokizane[5], Eiji Hase[5], Takeo Minamikawa[5,6], and Takeshi Yasui[5]

[1]Graduate School of Advanced Technology and Science, Tokushima University, 2-1 Minami-Josanjima, Tokushima, Tokushima 770-8506, Japan

[2]Graduate School of Sciences and Technology for Innovation, Tokushima University, 2-1 Minami-Josanjima, Tokushima, Tokushima 770-8506, Japan

[3]School of System Engineering, Kochi University of Technology, Kami, Kochi 782-8502, Japan

[4]Toho University, Funabashi, Chiba 274-8510, Japan

[5]Institute of Post-LED Photonics (pLED), Tokushima University, 2-1 Minami-Josanjima, Tokushima, Tokushima 770-8506, Japan

[6]Graduate School of Engineering Science, Osaka University, Toyonaka, Osaka 560-0043, Japan

*Both authors contributed equally to this work and are considered co-first authors

**Abstract**

Refractive-index (RI) sensing plays a pivotal role in various domains, encompassing applications like glucose sensing, biosensing, and gas detection. Despite the advantages of optical fiber sensors, such as their compact size, flexibility, and immunity to electromagnetic interference, they are often plagued by temperature-induced drift, which adversely impacts the accuracy of RI measurements. This study introduces an innovative approach to alleviate temperature-induced drift in RI-sensing optical frequency combs (OFCs) by employing active-dummy compensation. The central idea revolves around the utilization of a dual-comb setup, comprising an active-sensing OFC that monitors both sample RI and environmental temperature, and a dummy-sensing OFC that exclusively tracks environmental temperature. The disparity between these sensor signals, denoted as $\Delta f_{rep}$, effectively nullifies the effects of temperature variations, yielding a temperature-independent sensor signal for precise RI measurements. This investigation delves into the relationship between active-dummy temperature compensation and $\Delta f_{rep}$. It becomes evident that diminishing $\Delta f_{rep}$ values enhance temperature compensation, thereby diminishing fluctuations in $\Delta f_{rep}$ caused by environmental temperature shifts. This compensation technique establishes a direct link between $\Delta f_{rep}$ and sample RI, paving the way for absolute RI measurements based on $\Delta f_{rep}$. The findings of this research are a valuable contribution to the advancement of accurate and temperature-compensated RI sensing methodologies using dual-comb setup. The insights gained regarding $\Delta f_{rep}$ dependency and the strategies proposed for enhancing measurement precision and stability hold significant promise for applications in fields of product quality control and biosensing.

# 1. Introduction

Optical fiber sensor [1, 2] has been often used for refractive-index (RI) sensing and RI-sensing-based applications, including ethanol sensing [3], glucose sensing [4], DNA-interaction sensing [5], biomolecule detection [6], and antibody-antigen reaction [7]. RI fiber sensor benefits from inherent advantages of fiber sensor such as lightweight, small size, flexibility, cost effectiveness, immunity to electromagnetic interference, and environmental ruggedness. Various types of RI fiber sensors have been developed: tapered fiber [8], core-offset fiber [9], fiber Bragg grating (FBG) [10], surface plasmon resonance (SPR) [11], and multimode interference (MMI) [12]. Most of RI fiber sensors measures RI-dependent shift of optical spectrum peak or dip. However, the small shift to the relatively broad bandwidth in a spectral peak or dip as well as a limited spectrometer resolution of instruments often spoils the precision of those RI fiber sensors.

One potential approach to avoid the drawback of optical spectrum measurement in RI sensing is conversion of optical spectral sensor signal to photonic radio-frequency (RF) spectral sensor signal such as optical beat signal; namely, photonic RF conversion. The photonic RF conversion enables us to measure a sample RI as RF frequency signal, benefiting from high precision, wide dynamic range, convenience, and low cost in well - established electrical frequency measurements. One promising approach for photonic RF conversion is use of optical frequency comb (OFC) [13, 14] because OFC has a coherent link between optical frequency of OFC line ($\lambda$ m; typically, 194 THz in fiber OFC) and radio frequency of OFC line spacing ($f_{rep}$; typically, 100 MHz in fiber OFC) given by

$$\nu_m = f_{ceo} + m f_{rep} \qquad (1)$$

where $f_{ceo}$ is a carrier-envelope-offset frequency and m is a line number. Equation (1) indicates that an optical frequency signal of THz order can be uniquely converted to an electrical frequency signal of MHz order. Recently, fiber-OFC-based photonic RF conversion was adopted for fiber sensing of sample RI [15-17]. A fiber OFC including an intracavity MMI fiber sensor has been used for a photonic RF conversion of a sample RI into $f_{rep}$ via a combination of RI-dependent tunable bandpass filtering by the intracavity MMI fiber sensor and the wavelength dispersion by the OFC cavity fiber. Such RI-sensing OFC enables $f_{rep}$-reading RI sensing; due to the ultranarrow linewidth in the mode-locking oscillation, the $f_{rep}$ signal with the spectral linewidth below 1 Hz is precisely measured by a RF frequency counter synchronized with a rubidium frequency standard. Furthermore, the intracavity fiber sensor enables multiple interactions between the sample and the light, enhancing the sensitivity. However, the temperature drift of $f_{rep}$ caused by the temperature fluctuation of an optical cavity length (nL) spoils the reproducibility of $f_{rep}$-reading RI sensing because $f_{rep}$ in the RI-sensing OFC is free-running operation in contrast to actively stabilized operation of usual OFCs. Because of lack of one-to-one correspondence between $f_{rep}$ and sample RI, $f_{rep}$ change ($\Delta f_{rep}$) measurement is performed to obtain RI change. However, such relative measurement of $f_{rep}$ is not realistic in practical applications of RI sensing. If cavity-temperature-dependent slow drift of $f_{rep}$ is suppressed, the absolute measurement of the sample RI will be achieved based on the one-to-one correspondence between $f_{rep}$ and sample RI.

Recently, the active - dummy temperature - drift compensation with a dual - comb configuration was adopted for RI-sensing OFC and its application for biosensing [18]. A pair of RI-sensing OFCs was used for an active-sensing OFC and a dummy-sensing OFC. Since the active-sensing OFC provides the sensor signal ($f_{rep1}$) reflecting both the sample RI and the cavity temperature whereas the dummy-sensing OFC gives the sensor signal ($f_{rep2}$) reflecting only the cavity temperature, difference of sensor signals between them ($\Delta f_{rep}$) cancels the influence of temperature. In other words, it gives the temperature-independent sensor signal of sample RI. This dual-comb RI sensing enables the small change in the sensor signal caused by sample RI to be extracted from the large, variable background signal caused by temperature disturbance. Furthermore, it was effectively applied for rapid biosensing of severe acute respiratory syndrome coronavirus 2 (SARS-CoV-2) by help of antigen–antibody interactions [18]. However, an investigation regarding the optimal $\Delta f_{rep}$ in the context of the active-dummy temperature compensation and the one-to-one correspondence between sensor signal of $\Delta f_{rep}$ and sample RI has not yet been conducted.

In this article, we investigate the $\Delta f_{rep}$ dependence of the active-dummy temperature compensation, and then the $\Delta f_{rep}$ dependence of the one-to-one correspondence between sensor signals and sample RI by using as an ethanol water solution with different RIs as a sample.

## 2. Principle of operation

Figure 1(a) shows the principle of operation for RI-sensing OFC. Since its detailed principle is given elsewhere [15-17], we briefly described it. The first key element is an intracavity MMI fiber sensor [12], functioning as an RI-dependent tunable optical bandpass filter inside the cavity. Figure 1(b) shows a schematic diagram of the MMI fiber sensor. The MMI fiber sensor is composed of a clad-less multimode fiber (MMF; Thorlabs Inc., Newton, NJ, USA, FG125LA, core diameter = 125 µm, fiber

length = 58.94 mm) with a pair of single-mode fibers (SMFs) at both ends (Corning Inc., Corning, NY, USA, SMF28e +, core diameter = 8.2 μm, cladding diameter = 125 μm, fiber length = 150 mm). Only the exposed core of the clad-less MMF functions as a sensing part. The OFC light passing through the input SMF is diffracted at the entrance face of the clad-less MMF and then undergoes repeated total internal reflection at the boundary between the clad-less MMF core surface and the sample solution. As a result of constructive interference at the exit face of the clad-less MMF, only the light satisfying the MMI wavelength $\lambda_{MMI}$ can exit through the clad-less MMF and then be transmitted through the output SMF. $\lambda_{MMI}$ is given by

$$\lambda_{MMI} = \frac{n_{MMF} m_{MMI}}{L_{MMF}} [D(n_{sam})]^2 \qquad (2)$$

where $L_{MMF}$ and $n_{MMF}$ are the geometrical length and RI of the clad-less MMF, $m_{MMI}$ is the order of the MMI, $n_{sam}$ is the RI near the clad-less MMF core surface (namely, sample RI), and $D(n_{sam})$ is the effective core diameter of the clad-less MMF. Since $D(n_{sam})$ is influenced by the Goos-Hänchen shift on the core surface of the clad-less MMF, $\lambda_{MMI}$ is a function of the sample RI near the sensor surface. The intracavity MMI fiber sensor in this study functions as an RI-dependent optical bandpass filter tunable around $\lambda_{MMI}$ (= 1556.6 nm) with constructive interference at $L_{MMF}$ = 58.94 mm and $m_{MMI}$ = 4. This $\lambda_{MMI}$ was selected to match a spectral peak of the fiber OFC, suppressing the power loss. This intracavity MMI fiber sensor leads to RI-dependent optical spectrum shift of RI-sensing OFC (center wavelength = $\lambda_{MMI}$). The second key element is wavelength dispersion of refractive index in the OFC cavity fiber, converting RI-dependent $\lambda_{MMI}$ shift into RI-dependent shift of the optical cavity length nL. A relation between nL and $f_{rep}$ is given by

$$f_{rep} = \frac{c}{nL} \qquad (3)$$

where c is a velocity of light in vacuum, n and L are respectively a group refractive index and a geometrical length of fiber OFC cavity. Since $f_{rep}$ is a function of nL, RI-dependent nL shift is equivalent to RI-dependent $f_{rep}$ shift. Finally, one can read a sample RI as RI-dependent $f_{rep}$.

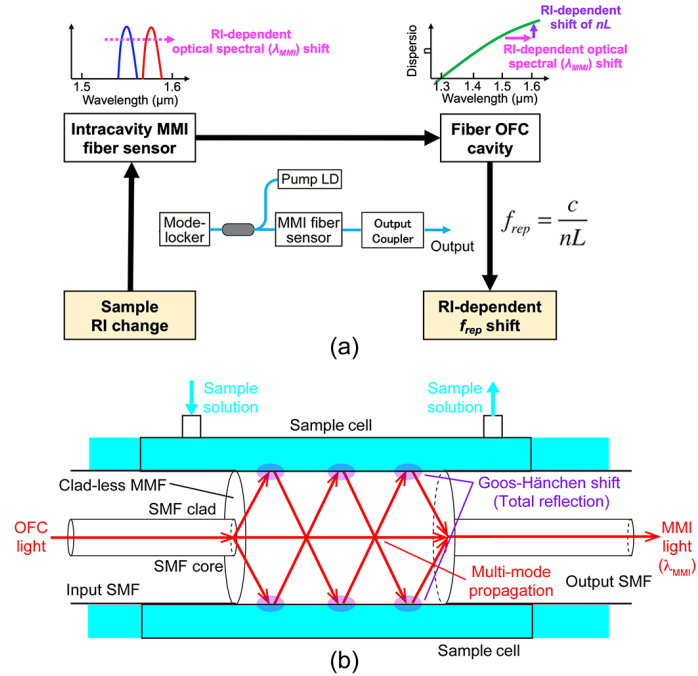

Fig. 1. (a) Principle of operation for RI sensing OFC. (b) Schematic drawing of MMI fiber sensor.

We next describe how to compensate the cavity-temperature-dependent drift of $f_{rep}$ in RI-sensing OFC. We here set a pair of RI-sensing OFCs with a sample and a reference material under equivalent environmental temperature disturbance as shown in Fig. 2(a). When the sample RI increases stepwise while the cavity temperature changes monotonously, the sensor signal $f_{rep1}$ of the active-sensing OFC changes depending on both sample RI and cavity temperature [blue plot in Fig. 2(b)]. On the other hand, when the cavity temperature change monotonously without the change of reference material RI in the dummy-sensing OFC, the sensor signal $f_{rep2}$ of the dummy-sensing OFC reflects only the cavity temperature. Therefore, a difference between $f_{rep1}$ and $f_{rep2}$ (= $\Delta f_{rep} = f_{rep1} - f_{rep2}$) reflects the sample RI signal without the influence of cavity temperature change [Fig. 2(c)].

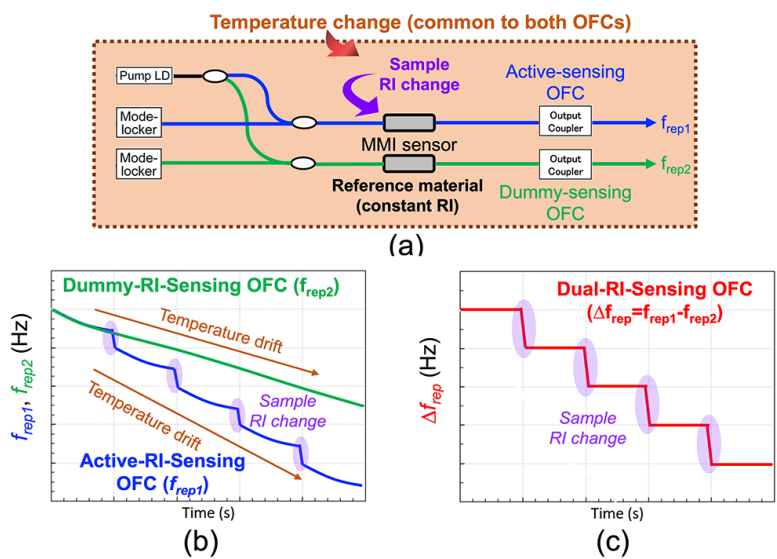

Fig. 2. Principle of operation for active-dummy temperature-drift compensation with dual-comb configuration in PR-sensing OFC. (a) Schematic drawing of dual-comb RI sensing setup. (b) Behavior of $f_{rep1}$ in active-sensing OFC and $f_{rep2}$ in dummy-sensing OFC. (c) Behavior of $\Delta f_{rep}$ in dual-comb RI sensing.

### 3. Experimental setup

Figure 3 shows the schematic drawing of the dual-comb RI sensing. They are based on a pair of linear fiber OFC cavities mode-locked by saturable absorption. Each linear cavity includes a polarization-maintaining single-mode fiber (PMF, Corning Panda PM1550, Corning, dispersion at 1550 nm = 17 ps•km$^{-1}$•nm$^{-1}$), an erbium-doped polarization-maintaining single-mode fiber (EDF, PM-ESF-7/125, Coherent, dispersion at 1550 nm = 16 ps•km$^{-1}$•nm$^{-1}$) [19], a fiber-coupled saturable absorbed mirror (SAM-1550-55-2ps-1.3b-0, BATOP, high reflection band = 1480-1640 nm, absorbance = 55 %, modulation depth = 34 %, relaxation time constant ~ 2 ps, size = 1.3-mm width, 1.3-mm height, 0.4-mm thickness), a wavelength-division-multiplexing coupler (WDM,

AFR, PMWDM-1-9801550-2-B-Q-6), a pumping laser diode (LD, BL976-PAG700, Thorlabs, wavelength = 976 nm, power = 700 mW), a 90:10 fiber-partial reflector (FPR, PMOFM-55-2-B-Q-F-90), and an intracavity MMI fiber sensor (MMI). The intracavity MMI fiber sensor was composed of a clad-less MMF (FG125LA, Thorlabs, core diameter= 125 μm, fiber length = 58. 94mm) with a pair of PMFs at both ends (core diameter=8.5μm, clad diameter=125μm). Here, we set $\lambda_{MMI}$ to 1556.6 nm. The fiber cavity was enclosed in a plastic box, and its temperature was controlled to 21.5 °C by a combination of a Peltier heater (TEC1-12708, Kaito Denshi, power = 76 W), a thermistor (NXFT15WF104FA2B050, Murata), and a temperature controller (TED200, Thorlabs, PID control) (not shown in Fig. 3). We fixed $f_{rep2}$ of the dummy-sensing OFC to be 31,451,121 Hz. Then, we adjusted $f_{rep1}$ of the active-sensing OFC around 31 MHz to change $\Delta f_{rep}$ within a range of a few kHz to 1,000 kHz by adjusting its cavity fiber length of the active-sensing OFC. We adopted mechanically sharing of linear fiber cavities to implement equivalent environmental temperature disturbance to the active-sensing and the dummy-sensing OFCs [20, 21], where all fiber components and intracavity MMI fiber sensors are placed at the same position of the cavity box. As a result, temperature drift and/or mechanical vibrations acting as disturbances on the cavity affect both cavities almost equivalently. The output light from them was detected by a pair of photodetectors (PD), and $f_{rep1}$ and $f_{rep2}$ were measured by an RF frequency counter (53220A, Keysight Technologies, frequency resolution = 12 digit•$s^{-1}$) synchronized to a rubidium frequency standard (FS725, Stanford Research Systems, accuracy = 5 × $10^{-11}$ and instability = 2 × $10^{-11}$ at 1s). Additionally, its optical spectrum was measured by an optical spectrum analyzer (AQ6315A, Yokogawa Electric Corp., wavelength accuracy = 0.02 nm, wavelength resolution = 0.02 nm).

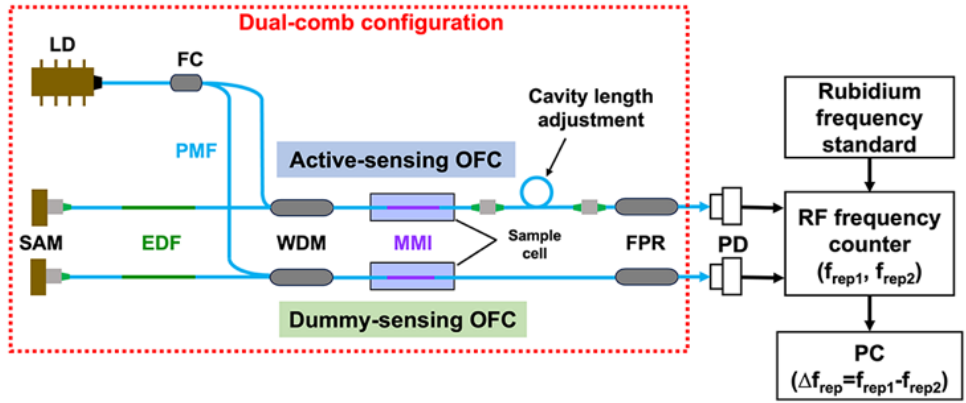

Fig. 3. Schematic drawing of experimental setup. LD, pumping laser diode; FC, 50:50 fiber coupler; PMF, polarization-maintaining single-mode fiber; EDF, erbium-doped polarization-maintaining single-mode fiber; SAM, fiber-coupled saturable absorbed mirror; WDM, wavelength-division-multiplexing coupler; MMI, intracavity multimode-interference fiber sensor; FPR, 90:10 fiber-partial reflector.

## 4. Results

- A. Basic performance

We first evaluated basic performance of active-sensing OFC and dummy-sensing OFC when a pure water was used as a sample for them. We prepared active-sensing OFCs with different $\Delta f_{rep}$ values (= 3 kHz, 46 kHz, 111 kHz, and 1,000 kHz). Figure 4(a) compares optical spectra of active-sensing OFC among different $\Delta f_{rep}$ values (resolution bandwidth = 0.1 nm). Each individual optical spectrum closely matches around the wavelength of 1559nm. For comparison, the optical spectrum of the dummy-sensing OFC was also measured as indicated by a black plot in Fig. 4(a). The overlap of optical spectra between the active-sensing OFC and the dummy-sensing OFC was not significant. The difference in overlap between them is due to the uncertainty in the clad-less MMF length ($L_{MMF}$) of the MMI fiber sensor and hence

MMI, which is caused by the fiber processing accuracy. Figures 4(b) and 4(c) show RF spectra of active-sensing OFC with $f_{rep1}$ and dummy-sensing OFC with $f_{rep2}$ when $\Delta f_{rep}$ was set to be 3 kHz (resolution bandwidth = 10 Hz). Similar RF spectra were observed for the active-sensing OFC with different $\Delta f_{rep}$ values (not shown). The narrow linewidth characteristics enable high precision of RI sensing, but they also make the system more susceptible to environmental temperature drift effects.

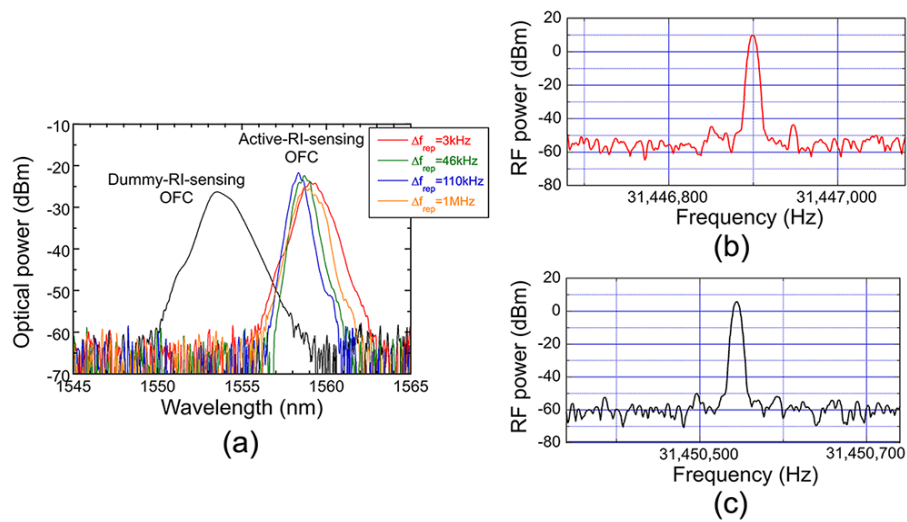

Fig. 4. Basic performance of active-sensing and dummy-sensing OFCs. (a) Comparison of optical spectra between active-sensing OFC with different $\Delta f_{rep}$ values and dummy-sensing OFC (RBW = 0.1 nm). RF spectra of (b) $f_{rep1}$ and (c) $f_{rep2}$ when $\Delta f_{rep}$ = 3 kHz (RBW = 10Hz).

We next evaluated the frequency instability of $f_{rep1}$ and $f_{rep2}$. Figure 5(a) shows a comparison of the frequency fluctuation between $f_{rep1}$ (red plots) and $f_{rep2}$ (black plots) with respect to gate time when $\Delta f_{rep}$ was set to be 3 kHz. Since both active-sensing OFC and dummy-sensing OFC have similar cavity configuration and environment, they show similar frequency instability of $f_{rep1}$ and $f_{rep2}$. A similar trend was observed for different $\Delta f_{rep}$ values (not shown), confirming that the frequency fluctuations of $f_{rep1}$ and $f_{rep2}$ are not dependent on $\Delta f_{rep}$. We also investigated the long-term frequency variations of $f_{rep1}$ in the active-sensing OFC. Figure 5(b) shows the temporal variations of $f_{rep1}$ within a range of 10 hours. We here defined the frequency deviation of $f_{rep1}$ from its initial value of measurement start as $\delta f_{rep1}$. Due to changes in the optical cavity length caused by variations in ambient temperature, fluctuations of $f_{rep1}$ within the 100 Hz range can be observed. The unidirectional increase in $f_{rep1}$ reflects gradual changes in the cavity environment temperature, likely due to ambient temperature variations. In contrast, the periodic repetitive changes correspond to room temperature fluctuations caused by air conditioning because the periodic fluctuations are synchronized with the simultaneous temperature measurements taken by the thermometer (not shown).

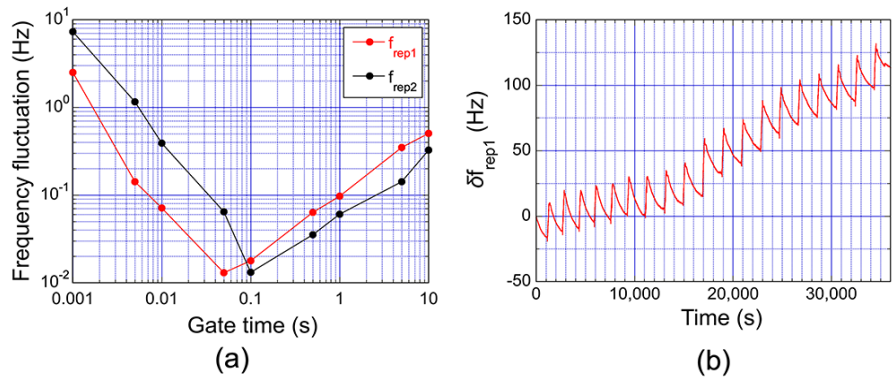

Fig. 5. Frequency characteristics of active-sensing and dummy-sensing OFCs. (a) Frequency fluctuation of $f_{rep1}$ and $f_{rep2}$ with respect to gate time. (b) Temporal variation of $f_{rep1}$ and $f_{rep2}$ in long term.

- B. Dependence of active-dummy temperature compensation on $\Delta f_{rep}$

As shown in Fig. 5, the active-sensing OFC and the dummy-sensing OFC exhibited similar behaviors in frequency characteristics, which enables compensation for the temperature drift effect by taking the difference between them. Of particular interest here is how this compensation effect depends on $\Delta f_{rep}$ and whether an optimal $\Delta f_{rep}$ exists. We consider the alignment of optical cavity lengths, given by an inverse of $f_{rep}$, in the active-sensing and dummy-sensing OFCs to be an important parameter in the active-dummy compensation using mechanically-sharing dual-comb configuration. Therefore, an investigation was conducted regarding the dependence of active-dummy temperature compensation on $\Delta f_{rep}$ (= 3 kHz, 46 kHz, 111 kHz, and 1,000 kHz); these four $\Delta f_{rep}$ values correspond to alignments of 99.99%, 99.87%, 99.52%, and 96.82%, respectively, aiming to evaluate the extent of influence from alignments with different orders of magnitude. Here, considering the results indicating changes of approximately 100Hz in $f_{rep1}$ due to room temperature variations as shown in Fig. 5(b), we measured $\Delta f_{rep}$ when $f_{rep1}$ was changed by approximately 100 Hz by rapid air conditioning. We used a pure water as a sample with a constant RI for them, again. We measured the temporal variations of $f_{rep1}$, $f_{rep2}$, and $\Delta f_{rep}$ when pure water was used as a sample for both OFCs. Figure 6 shows the temporal shifts in $f_{rep1}$, $f_{rep2}$, and $\Delta f_{rep}$, namely, $\delta f_{rep1}$, $\delta f_{rep2}$, and $\delta \Delta f_{rep}$, respectively, when $\Delta f_{rep}$ was set to be (a) 3 kHz, (b) 46 kHz, (c) 111 kHz, and (d) 1000 kHz. Comparing with $\delta f_{rep1}$ and $\delta f_{rep2}$, it is evident that the width of the frequency fluctuations in $\Delta f_{rep}$ has been significantly reduced, indicating the effective operation of the active-dummy temperature compensation. Furthermore, the width of these fluctuations decreases as $\Delta f_{rep}$ is reduced to lower frequencies. A small $\Delta f_{rep}$ implies that the difference in optical cavity length between the active-sensing OFC and the dummy-sensing OFC is small, indicating that the thermal changes in optical cavity length have a similar impact on both. This is believed to be the origin of this $\Delta f_{rep}$ dependence. Therefore, a smaller $\Delta f_{rep}$ is preferred, but there is a limit imposed by the machining precision of the cavity fiber length. For example, the uncertainty of 1 mm in optical cavity length leads to that of 10 kHz in $f_{rep1}$. The actual machining precision of the fiber cutter is around 1 mm, making it difficult to adjust $\Delta f_{rep}$ below 1 kHz.

We also investigated the dependence of its frequency instability on $\Delta f_{rep}$. Figure 6(e) compares the frequency fluctuation among different $\Delta f_{rep}$ values with respect to gate time. In contrast to Figs. 6(a), 6(b), 6(c), and 6(d), no clear $\Delta f_{rep}$ dependence was observed. This frequency instability reflects short-term rather than long-term behavior, and it was found that $\Delta f_{rep}$ dependence does not appear on such short time scales. From the viewpoint of RI sensing, the frequency instability in long-term is important.

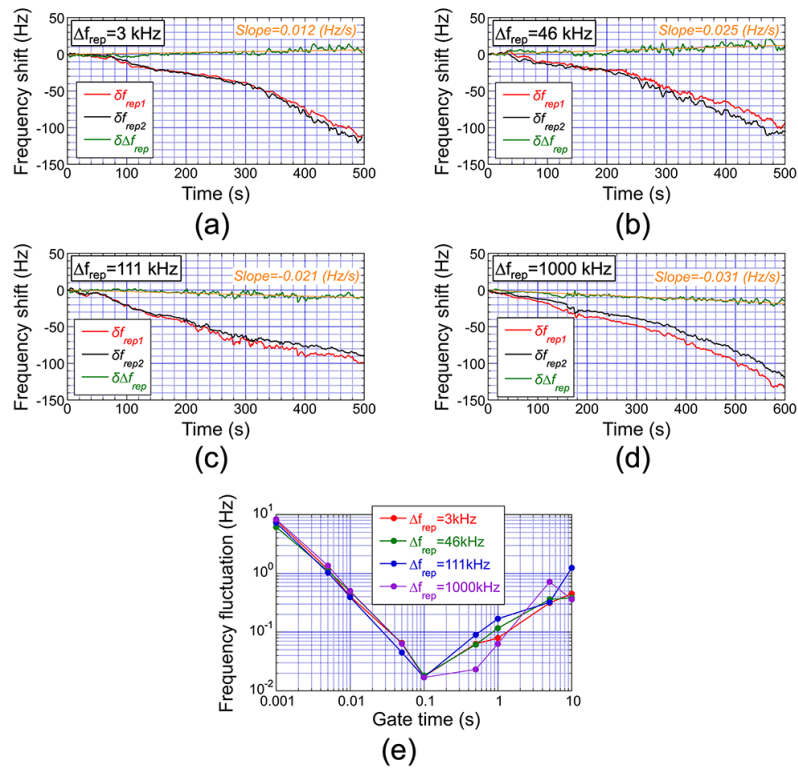

Fig. 6. Temporal shifts in $f_{rep1}$, $f_{rep2}$, and $\Delta f_{rep}$ ($\delta f_{rep1}$, $\delta f_{rep2}$, and $\delta\Delta f_{rep}$) when $\Delta f_{rep}$ was set to be (a) 3 kHz, (b) 46 kHz, (c) 111 kHz, and (d) 1000 kHz. (e) Comparison of frequency fluctuation among different $\Delta f_{rep}$ values with respect to gate time.

- C. Dependence of temperature-compensated RI sensing performance on $\Delta f_{rep}$

Finally, we investigated the dependence of active-dummy temperature-compensated RI sensing performance on $\Delta f_{rep}$ (= 3 kHz, 40 kHz, 150 kHz, and 1,000 kHz). We used ethanol solutions consisting of water (RI = 1.3180 refractive index unit or RIU at 1550 nm) and ethanol (RI = 1.347 RIU at 1550 nm) at different ratios, corresponding to different RIs, as target samples in the active-sensing OFC. We prepared six samples with different RIs (= 0 EtOH%, 2.5 EtOH%, 5 EtOH%, 7.5 EtOH%, 10 EtOH%, corresponding to 1.3180 RIU, 1.3188 RIU, 1.3196 RIU, 1.3205 RIU, and 1.3213 RIU). Additionally, pure water (a 0 vol% ethanol solution, corresponding to 1.3180 RIU) was used as a reference material with a constant RI in the dummy-sensing OFC. We repeated the RI sensing of ethanol solutions with different concentrations five times based on the measurement of $f_{rep1}$. Figure 7 shows the average and standard deviation of $f_{rep1}$ measured with different RI samples when $\Delta f_{rep}$ was set to be (a) 3 kHz, (b) 40 kHz, (c) 150 kHz, and (d) 1000 kHz. While a relatively large standard deviation is observed, the standard deviation varies significantly with $\Delta f_{rep}$. The reason for this behavior, which should ideally show similar fluctuations in $f_{rep1}$ regardless of the value of $\Delta f_{rep}$, is believed to be due to the behavior of temperature drift, as shown in Fig. 5(b). Depending on the timing of the measurement start, $f_{rep1}$ is influenced differently by temperature drift. This results in the difficulty to establish the one-to-one correspondence between $f_{rep1}$ and sample RI.

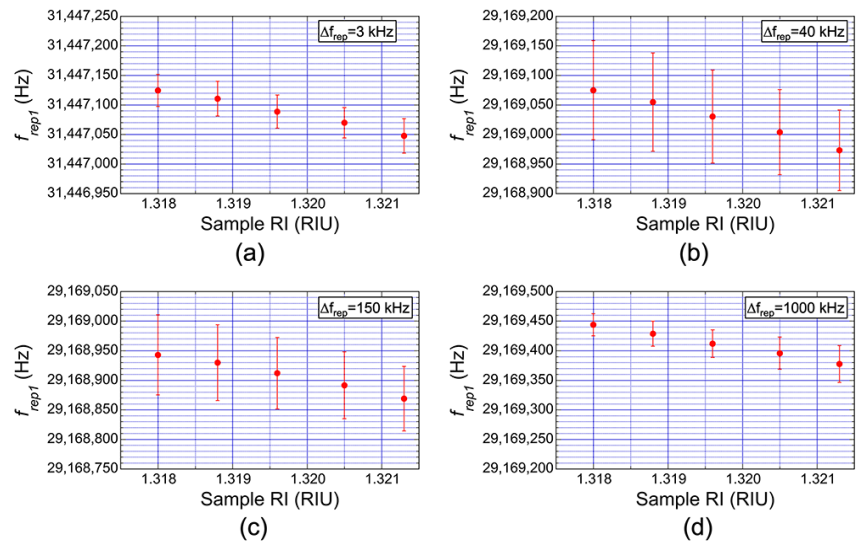

Fig. 7. Average and standard deviation of $f_{rep1}$ measured with different RI samples: (a) $\Delta f_{rep}$ = 3 kHz, (b) $\Delta f_{rep}$ = 40 kHz, (c) $\Delta f_{rep}$ = 150 kHz, and (d) $\Delta f_{rep}$ = 1000 kHz.

We next repeated the RI sensing of ethanol solutions with different concentrations five times based on the measurement of $\Delta f_{rep}$. Figure 8 shows the average and standard deviation of $\Delta f_{rep}$ measured with different RI samples when $\Delta f_{rep}$ was set to be (a) 3 kHz, (b) 40 kHz, (c) 150 kHz, and (d) 1000 kHz. For instance, when $\Delta f_{rep}$ is 3 kHz, 46 kHz, and 111 kHz, the standard deviation of $\Delta f_{rep}$ (typically, a few Hz) is significantly smaller than that of $f_{rep1}$ [typically, several tens Hz; see Figs. 7(a), 7(b), and 7(c)]. This indicates that the active-dummy temperature compensation is functioning, and a one-to-one correspondence between $\Delta f_{rep}$ and sample RI is established independently of environmental temperature changes. This suggests the potential for RI sensing based on the absolute measurement of $\Delta f_{rep}$. However, when $\Delta f_{rep}$ is 1,000 kHz, the standard deviation of $f_{rep1}$ is equivalent to that of $\Delta f_{rep}$. In this case, active-dummy temperature compensation is not functioning. This aligns significantly with the results in Fig. 6(d). Thus, we confirmed the dependence of the active-dummy temperature-compensated RI sensing performance on $\Delta f_{rep}$. Since the detailed setting of $\Delta f_{rep}$ is challenging to achieve below 10 kHz due to the precision of the cavity fiber length machining, it is preferable to choose an optimal value for $\Delta f_{rep}$ that is 110 kHz or lower, providing some flexibility in the selection process.

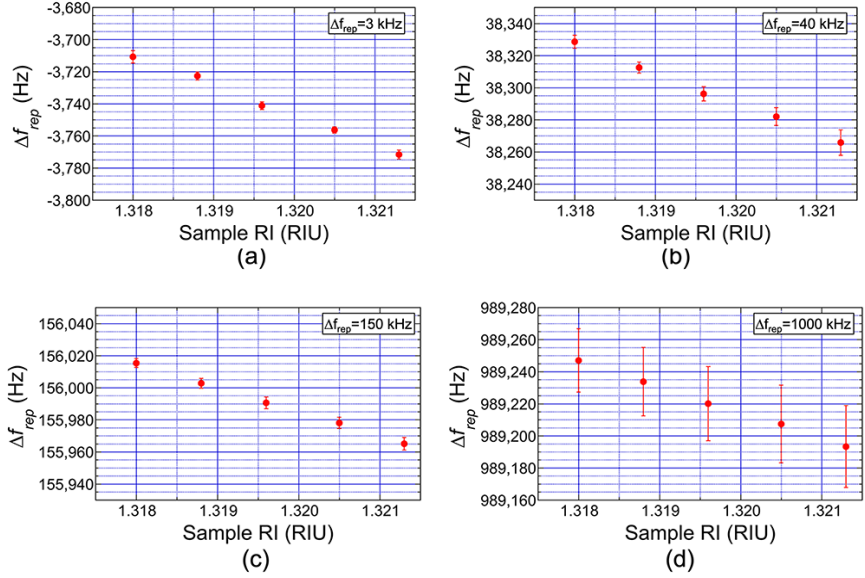

Fig. 8. Average and standard deviation of Δf$_{rep}$ measured with different RI samples: (a) Δf$_{rep}$ = 3 kHz, (b) Δf$_{rep}$ = 40 kHz, (c) Δf$_{rep}$ = 150 kHz, and (d) Δf$_{rep}$ = 1000 kHz.

## 5. Discussion

In the active-dummy temperature-compensated RI sensing using a mechanically sharing dual-cavity dual-comb configuration, we have successfully suppressed temperature drift. The merit of this active-dummy temperature compensation is expected to be highlighted in improving measurement reproducibility of RI sensing. Unlike conventional RI-sensing OFCs based on f$_{rep}$ measurements, which suffer from poor measurement reproducibility due to temperature drift, the suppression of temperature drift by active-dummy temperature control allows for the establishment of a good one-to-one correspondence between sample RI and Δf$_{rep}$. To succinctly illustrate this point with data, we conducted measurements on five sets of water/ethanol solution samples with varying RIU. Figure 9(a) shows the relationship between sample RI and sensor signal when the f$_{rep1}$ (≈ 31.4 MHz) was used as the measurement signal. It can be observed that their resulting linear slopes deviate significantly with each measurement due to temperature drift, which fluctuates constantly as a background offset. This deviation compromises the one-to-one correspondence between sample RI and sensor signal, leading to reduced measurement reproducibility. On the other hand, Figure 9(b) depicts the relationship between sample RI and sensor signal when the $\Delta f_{rep}$ (≈ 3.7 kHz) was used as the measurement signal. It is evident that the resulting linear slopes obtained from five repeated measurements are well-aligned, indicating effective suppression of temperature drift by active-dummy temperature compensation. In this way, the active-dummy temperature compensation allows for absolute RI measurements based on the one-to-one correspondence between $\Delta f_{rep}$ and sample RI.

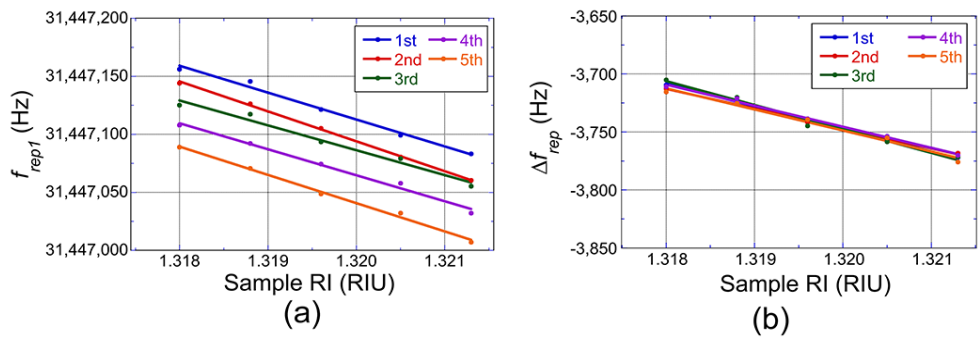

Fig. 9. Relationship between sample RI and sensor signal when r . (a) Use of $f_{rep1}$ (≈ 31.4 MHz) as a measurement signal. (b) Use of $\Delta f_{rep}$ (≈ 3.7 kHz) as a measurement signal.

However, as shown in Fig. 6, there is still some residual fluctuation in $\Delta f_{rep}$ that could be further reduced, and here we discuss several challenges for achieving that.

**(a) Lower $\Delta f_{rep}$ values:** As shown in Fig. 6, reducing $\Delta f_{rep}$ values improves temperature compensation. However, fine adjustments of $\Delta f_{rep}$ are limited to a few kHz due to the precision of cavity fiber length machining. Achieving even lower $\Delta f_{rep}$ values may be challenging as long as two independent fiber cavities are used. Inserting an optical path length adjustment mechanism using free-space optics into the cavity enables further fine adjustments; however, it may compromise the active-dummy temperature compensation effect.

**(b) Spectral overlap of active-sensing and dummy-sensing OFCs:** To ensure that the temperature drift of $f_{rep1}$ and $f_{rep2}$ due to thermal changes in the optical cavity length appears similarly in both OFCs, it is essential to have well-overlapped spectra. In the present setup, as shown in Fig. 4(a), the spectra of both OFCs did not match closely. This discrepancy could be attributed to slight differences in the characteristics, namely MMI, of the intracavity MMI fiber sensors, likely due to variations in fiber length machining precision. As shown in Eq. (2), $\lambda_{MMI}$ is given by the inverse of $L_{MMF}$. Typically, the fiber length error from our fiber processing equipment is around 1 mm. If $L_{MMF}$ is 58 mm, this introduces an uncertainty of approximately 1.8%. This uncertainty is directly reflected in MMI, potentially resulting in an error within 28 nm at MMI of 1556.6 nm. The discrepancy in the optical spectra between the dummy-RI-sensing OFC and the active-RI-sensing OFC in Fig. 4(a) accurately reflects this situation. Achieving further spectral overlap may also be challenging as long as two independent intracavity MMI fiber sensors are used.

**(c) Precise alignment of cavity fiber configuration:** Another consideration is how accurately the cavity fiber and components can be placed to ensure that both OFCs experience equivalent temperature disturbances. However, there are limits to this alignment, as adjustments need to be made within the confined space of the cavity box.

As such, these three challenges may be challenging with the current experimental setup configuration.

To potentially overcome these challenges, one approach could be the use of a single-cavity dual-OFC configuration. Recently, single-cavity dual-comb fiber lasers have gained attention as light sources for dual-comb spectroscopy [22-24]. These fiber lasers achieve dual-comb mode-locked oscillation in a single cavity, resulting in reduced $\Delta f_{rep}$ fluctuations to below 0.1 Hz without the need for active laser control. This approach has a potential to overcome three challenges above (lower $\Delta f_{rep}$ values, spectral overlap of active-sensing and dummy-sensing OFCs, and precise alignment of fiber configuration). The key point is that by multiplexing mode-locked oscillation, the same fiber OFC cavity is shared between the two OFCs, making disturbances such as temperature drift and mechanical vibrations completely common to each other. As a result, it becomes possible to maximize the effectiveness of active-dummy compensation. Integrating an MMI fiber sensor into these lasers, including the possibility of sharing even an intracavity MMI fiber sensor, is technically challenging. However, this combination has the potential to significantly improve the performance of dual-comb RI sensing while simplifying the experimental setup. Please note that further exploration and experimentation may be required to determine the feasibility and effectiveness of these approaches.

## 6. Conclusion

In this article, we explored the use of dual OFC configuration for RI sensing and investigated the dependence of active-dummy temperature compensation and its relationship with RI sensing on $\Delta f_{rep}$. The active-dummy temperature compensation technique, involving taking the difference between sensor signals from the active-sensing and dummy-sensing OFCs, effectively reduced the influence of temperature drift of sensor signal on RI sensing, allowing for more accurate measurements of sample RI. Regarding the dependence of the effectiveness of temperature compensation on $\Delta f_{rep}$, smaller $\Delta f_{rep}$ values led to improved temperature compensation, resulting in reduced fluctuations in $\Delta f_{rep}$ due to environmental temperature changes. In the temperature-compensated RI sensing, we demonstrated that active-dummy temperature compensation, particularly at lower $\Delta f_{rep}$ values, enabled a one-to-one correspondence between $\Delta f_{rep}$ and sample RI, suggesting the potential for absolute RI measurements based on $\Delta f_{rep}$. We also discussed a potential avenue to overcome three challenges for further enhanced performance in dual-comb RI sensing.

This research contributes to the development of precise and temperature-compensated RI sensing techniques using dual-comb OFCs. It provides insights into the dependence of temperature compensation on $\Delta f_{rep}$ and suggests strategies for further improving the accuracy and stability of RI measurements in various applications, including quality monitoring of industrial products and biosensing.

**Acknowledgments**

The authors acknowledge Prof. Kaoru Minoshima at The University of Electro-Communications for her help in the mechanical sharing dual-comb configuration.